\documentclass[twocolumn,superscriptaddress,showpacs,preprintnumbers,amsmath,amssymb]{revtex4}
\usepackage{amsmath}
\usepackage{amssymb}
\usepackage{dcolumn}
\usepackage{graphicx}
\usepackage{bm}
\usepackage{epstopdf}
\usepackage{threeparttable}
\usepackage{float}
\usepackage{color}

\usepackage [english]{babel}
\usepackage [autostyle, english = american]{csquotes}
\MakeOuterQuote{"}

\makeatletter
\newcommand*{\rom}[1]{\expandafter\@slowromancap\romannumeral #1@}
\makeatother
\makeatletter
\def\p@subsection{}
\makeatother

\begin{document}
\title{The effects of cooling rate on particle rearrangement statistics: 
Rapidly cooled glasses are more ductile and less reversible}

\author{Meng Fan} 
\affiliation{Department of Mechanical Engineering and Materials Science, Yale University, New Haven, Connecticut, 06520, USA}
\affiliation{Center for Research on Interface Structures and Phenomena, Yale University, New Haven, Connecticut, 06520, USA}

\author{Minglei Wang} 
\affiliation{Department of Mechanical Engineering and Materials Science, Yale University, New Haven, Connecticut, 06520, USA}
\affiliation{Center for Research on Interface Structures and Phenomena, Yale University, New Haven, Connecticut, 06520, USA}

\author{Kai Zhang} 
\affiliation{Department of Chemical Engineering, Columbia University, New York, New York 10027, USA}

\author{Yanhui Liu} 
\affiliation{Department of Mechanical Engineering and Materials Science, Yale University, New Haven, Connecticut, 06520, USA}
\affiliation{Center for Research on Interface Structures and Phenomena, Yale University, New Haven, Connecticut, 06520, USA}
\author{Jan Schroers}
\affiliation{Department of Mechanical Engineering and Materials Science, Yale University, New Haven, Connecticut, 06520, USA}
\affiliation{Center for Research on Interface Structures and Phenomena, Yale University, New Haven, Connecticut, 06520, USA}
\author{Mark D. Shattuck}
\affiliation{Department of Physics and Benjamin Levich Institute, The City College of the City University of New York, New York, New York, 10031, USA}
\affiliation{Department of Mechanical Engineering and Materials Science, Yale University, New Haven, Connecticut, 06520, USA}
\author{Corey S. O'Hern}
\affiliation{Department of Mechanical Engineering and Materials Science, Yale University, New Haven, Connecticut, 06520, USA}
\affiliation{Center for Research on Interface Structures and Phenomena, Yale University, New Haven, Connecticut, 06520, USA}
\affiliation{Department of Physics, Yale University, New Haven, Connecticut, 06520, USA}
\affiliation{Department of Applied Physics, Yale University, New Haven, Connecticut, 06520, USA}

\date{\today}

\begin{abstract}
Amorphous solids, such as metallic, polymeric, and colloidal glasses,
display complex spatiotemporal response to applied deformations.  In
contrast to crystalline solids, during loading, amorphous solids
exhibit a smooth crossover from elastic response to plastic flow.  In
this study, we investigate the mechanical response of binary
Lennard-Jones glasses to athermal, quasistatic pure shear as a
function of the cooling rate used to prepare them.  We find several
key results concerning the connection between strain-induced particle
rearrangements and mechanical response. We show that more rapidly
cooled glasses undergo more frequent and larger particle
rearrangements than slowly cooled glasses.  We find that the ratio of
the shear to bulk moduli decreases with increasing cooling rate, which
suggests that more rapidly cooled glasses are more ductile than slowly
cooled samples. In addition, we characterized the degree of
reversibility of particle motion during cyclic shear.  We find that
irreversible particle motion occurs even in the putative linear regime
of stress versus strain.  However, slowly cooled glasses, which
undergo smaller rearrangements, are more reversible under cyclic shear
than rapidly cooled glasses.  Thus, we show that more ductile glasses
are also less reversible.

\end{abstract}

\pacs{62.20.-x,
63.50.Lm
64.70.kj
64.70.pe
} 

\maketitle

\begin{figure}
\begin{center}
\includegraphics[width=1.00\columnwidth]{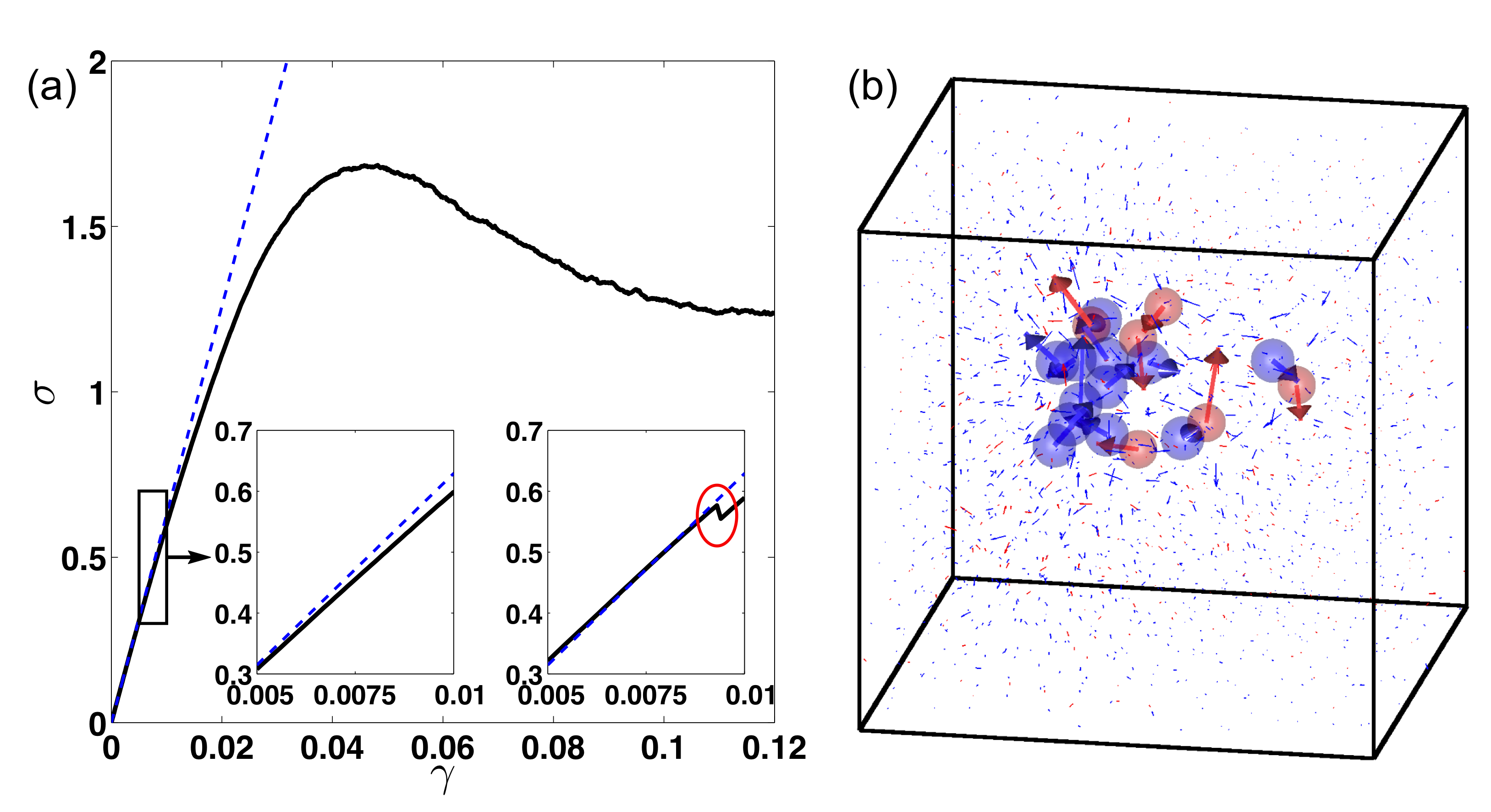}
\caption{(a) von Mises stress $\sigma$ versus strain $\gamma$ (solid
line) from simulations of a binary Lennard-Jones glass (prepared at
cooling rate $R_c =10^{-3}$) with $N=2000$ spheres undergoing
athermal, quasistatic pure shear averaged over $500$ samples.
Periodic boundary conditions are employed in the simulations, which
prevent fracture during loading. The affine stress versus strain
obtained in the $\gamma \rightarrow 0$ limit is given by the dashed
line. The left inset provides a close-up of the ensemble-averaged
$\sigma(\gamma)$, which highlights the deviation from linear
behavior in the range $\gamma=0.005$ to $0.01$. The right inset
gives $\sigma(\gamma)$ for a single sample over the same small
strain interval.  The circled stress drop indicates the particle
rearrangement event in (b). The vectors (which have been scaled by a
factor of $15$) indicate the particle displacements that caused the
stress drop. The participation
number~\cite{malandro1999relationships} of this event is roughly
$18$. Blue and red spheres indicate the large and small particles,
respectively, with the largest displacements. }
\label{fig:event}
\end{center}
\end{figure}

Amorphous solids, including metallic, polymeric, and colloidal
glasses, possess complex mechanical response to applied deformations,
such as plastic flow~\cite{falk1998dynamics,utz2000atomistic,
  maloney2004subextensive,sun2010plasticity}, strain
localization~\cite{spaepen_1977,schall2007structural,ju2011atomically,ding2014soft,
  jensen2014local}, creep flow~\cite{ju2011atomically,Atzmon_2014,Ju_2015}, and fracture~\cite{chen_2016,antonaglia2014tuned,kumar2013critical}. In crystalline
materials, topological defects reflecting the symmetry of the
crystalline phase govern the response to applied deformations. In
amorphous solids without long-range positional order, it is more
difficult to detect and predict changes from elastic response to
irreversible behavior~\cite{ding2014soft,gendelman2015shear}, such as
yielding~\cite{park2008elastostatically,regev2015reversibility} and
flow~\cite{harmon2007anelastic,sun2010plasticity}. The typical
response of the deviatoric stress to an applied (pure) shear strain
for amorphous solids is depicted in Fig.~\ref{fig:event} (a). The
average stress increases roughly linearly with strain for small
strains, indicating a putative elastic regime.  At larger strains, the
stress response softens and becomes anelastic, but it continues to
increase with strain.  For larger strains ({\it i.e.}  near $\gamma
\sim 0.05$), the shear stress reaches a peak (whose height depends on
the thermal history of the glass) and then begins to decrease until it
plateaus at a steady state value in the plastic flow regime~\cite
{utz2000atomistic,harmon2007anelastic}. (For this system, we employed 
boundary conditions that prevent fracture.)

Several recent studies have suggested that amorphous solids do not
possess a truly elastic response
regime~\cite{park2008elastostatically,
  fujita2012low,dmowski2010elastic,schall2007structural,jensen2014local,
  ju2011atomically,ding2012correlating}.  For example, both a
sublinear increase of the average stress with strain (left inset to
Fig.~\ref{fig:event} (a)) and rapid drops in stress over narrow strain
intervals (right inset to Fig.~\ref{fig:event} (a)) have been observed
at strains below the nominal yield strain of
$2\%$~\cite{ding2012correlating,park2008elastostatically,fujita2012low}.
The rapid drops in stress are caused by particle rearrangements ({\it
  e.g.}  in Fig.~\ref{fig:event} (b)), which are often referred to as
shear transformation zones~\cite
{argon1979plastic,falk1998dynamics,langer2008shear}.  We will show
below how the frequency and size of particle rearrangements determine
the mechanical response of amorphous solids.

In this article, we build a conceptual framework for the mechanical
response of amorphous solids by exploring the potential energy
landscape and particle rearrangement statistics of binary
Lennard-Jones glasses during athermal, quasistatic pure shear.  The
initial glasses are prepared over a wide range of cooling rates. The
cooling rate determines the fictive temperature, which defines the
average energy of the glass in the potential energy
landscape~\cite{debenedetti2001supercooled}. The fictive temperature
significantly affects mechanical properties, such as ductility~\cite
{kumar2013critical,kumar2011unusual}, shear band formation~\cite
{zemp2015crystal}, and the stress-strain
relation~\cite{utz2000atomistic,ashwin2013cooling}.  Our key result is
that more rapidly cooled glasses possess more frequent and larger
potential energy drops during applied shear strain compared to more
slowly cooled glasses.  As a result, the ensemble-averaged (metabasin)
curvature of the energy landscape is much smaller for rapidly cooled
glasses.  We connect the statistics of particle rearrangements to
whether amorphous solids exhibit brittle or ductile mechanical
response and characterize the degree of irreversibility of particle
rearrangements in response to shear reversal. We find that more
rapidly cooled glasses are more ductile and irreversible compared to
slowly cooled glasses.

We performed constant number, pressure, and temperature (NPT)
molecular dynamics (MD) simulations of binary Lennard-Jones mixtures
containing $80\%$ large and $20\%$ small spherical particles by number
(both with mass $m$) in a cubic box with volume $V$ and periodic
boundary conditions.  The particles interact pairwise via the
Kob-Andersen, shifted-force potential, $u(r_{ij}) = 4
\epsilon_{ij}[(\sigma_{ij}/r_{ij})^{12}-(\sigma_{ij}/r_{ij})^6]$,
where $r_{ij}$ is the separation between particles $i$ and $j$,
$u(r_{ij})= 0$ for $r_{ij} > 2.5 \sigma_{ij}$, and the energy and
length parameters are given by $\epsilon_{AA}=1.0$,
$\epsilon_{BB}=0.5$, $\epsilon_{AB}=1.5$, $\sigma_{AA}=1.0$,
$\sigma_{BB}=0.88$, and $\sigma_{AB}=0.8$~\cite{kob1995testing}.
Energy, temperature, pressure, and time scales are expressed in units of
$\epsilon_{AA}$, $\epsilon_{AA}/k_B$, $\epsilon_{AA}/\sigma_{AA}^3$,
and $\sigma_{AA} \sqrt{m/\epsilon_{AA}}$, respectively, where $k_B$ is
Boltzmann's constant~\cite{allen1989computer}.

We first equilibrate systems in the liquid regime at constant
temperature $T_0 = 0.6$ and pressure $P=0.025$ using a Nos\'{e}-Hoover
thermostat and barostat, a second-order simplectic integration
scheme~\cite{plimpton1995fast,tuckerman2006liouville}, and time step
$\Delta t = 10^{-3}$.  We cool systems into a glassy state at zero
temperature using a linear cooling ramp, $T(t) = T_0-R_c t$ over a
range of cooling rates from $R_c = 10^{-2}$ to $10^{-5}$, all of which
are above the critical cooling rate. Thus, all of the zero-temperature
samples are disordered. We then apply athermal, quasistatic pure shear
at fixed pressure.  (See an expanded discussion of the methods in
the Supplemental Material.) To do this, we expand the box length and
move all particles affinely in the $x$-direction by a small strain
increment $d\gamma_x=d\gamma=10^{-4}$ and compress the box length and
move all particles affinely in the $y$-direction by the same strain
increment $d\gamma_y=-d\gamma$.  Following the applied pure shear
strain, we minimize the total enthalpy of the system $H={\cal U} + PV$
at fixed pressure $P=10^{-8}$, where ${\cal U}$ is the total potential
energy. We successively apply pure strain increments $d\gamma$ and
minimize the enthalpy at fixed pressure after each increment to a
total strain $\gamma$. We studied systems with $N=250$, $500$, $1000$,
and $2000$ particles to assess finite size effects.

We developed a method to unambiguously determine if a
particle rearrangement occurs with an accuracy on the order of
numerical precision, which allows us to detect rearrangements with
sizes ranging over more than seven orders of magnitude.  To identify particle
rearrangement events, at each strain $\gamma$ we compare the total
potential energy per particle $U(\gamma)={\cal U}(\gamma)/N$ from
simulations undergoing forward shear to the potential energy per
particle $U'(\gamma)$ obtained by first a forward shear step from
strain $\gamma$ to $\gamma+d\gamma$ (and enthalpy minimization)
followed by a backward shear step from $\gamma+d\gamma$ back to
$\gamma$ (and enthalpy minimization). We find that the distribution of
the magnitudes of the energy differences $|\Delta U(\gamma)| =
|U(\gamma)-U'(\gamma)|$ is bimodal with peaks near $10^{-14}$
corresponding to numerical error and $10^{-3}$ corresponding to
distinct particle rearrangements. Thus, it is straightforward to
identify particle rearrangements as those with $|\Delta U| > U_{t}$,
where the threshold $U_t = 10^{-10}$ clearly distinguishes numerical
error from particle rearrangements.  (See Supplemental Material.) We
denote the total number of rearrangements in the strain interval $0$
to $\gamma$ as $N_r(\gamma)$. In addition, we calculate the total
energy lost after each rearrangement $i=1,\ldots,N_r(\gamma)$ over the
strain interval $\gamma$: $U_{\rm loss} = \sum_{i=1}^{N_r(\gamma)}
|\Delta U(\gamma_i)|$, where $\gamma_i$ indicate the strains at which
rearrangements occur.

\begin{figure}
\begin{center}
\includegraphics[width=0.9\columnwidth]{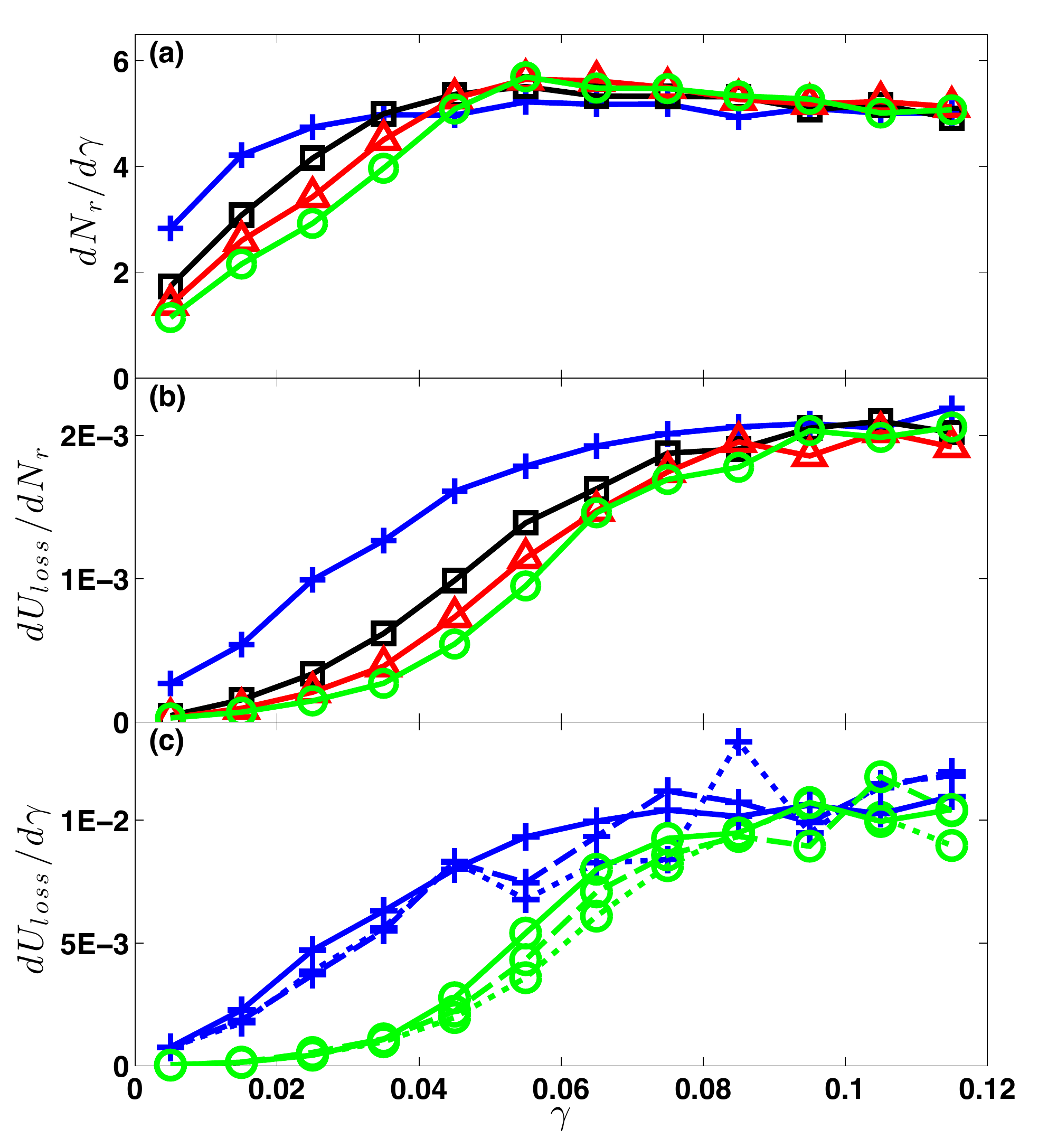}
\caption{(a) Particle rearrangement frequency $dN_r/d\gamma$, (b) mean
energy drop per rearrangement $dU_{\rm loss}/dN_r$, and (c) the product of (a) 
and (b), the
energy drop per strain $dU_{\rm loss}/d\gamma$, as a function of
strain for several cooling rates $R_c=10^{-2}$ (crosses), $10^{-3}$
(squares), $10^{-4}$ (triangles), and $10^{-5}$ (circles) used to
prepare the binary Lennard-Jones glasses.  In (a) and (b), the 
curves are obtained by
averaging over $500$ samples containing $N=2000$ particles. In (c), 
we show $dU_{\rm
loss}/d\gamma$ for three system sizes: $N=2000$ (solid lines),
$1000$ (dashed lines), and $500$ (dotted lines).}
\label{fig:cata}
\end{center}
\end{figure}  

In Fig.~\ref{fig:cata} (a) and (b), we plot the frequency of
rearrangements $dN_r/d\gamma$ and energy loss per rearrangement
$dU_{\rm loss}/dN_r$ as a function of strain.  Both the frequency and
energy loss increase with strain for small strains ($\gamma < 0.05$
for $dN_r/d\gamma$ and $\gamma < 0.08$ for $dU_{\rm loss}/dN_r$) and
then reach plateau values that are independent of strain.  Both
quantities are sensitive to cooling rate in the small strain regime:
glasses quenched using more rapid cooling rates ({\it i.e.} $R =
10^{-2}$) incur more and larger particle rearrangements. For more
slowly cooled glasses ({\it i.e.} $R=10^{-5}$), the systems only begin
losing energy (as measured from $dU_{\rm loss}/d\gamma$) beyond a
characteristic strain $\gamma_c \approx 0.02$.  In contrast, for
rapidly cooled glasses, the energy loss is roughly proportional to
strain for $\gamma < 0.06$.  At large strains $\gamma > 0.06$,
$dU_{\rm loss}/d\gamma$ becomes independent of cooling rate and
strain.  Further, we find that $d U_{\rm loss}/d\gamma$, which is the
product of $dN_r/d\gamma$ and $dU_{\rm loss}/dN_r$, is roughly
independent of system size over the range of $N$ we consider.

\begin{figure}
\begin{center}
\includegraphics[width=0.9\columnwidth]{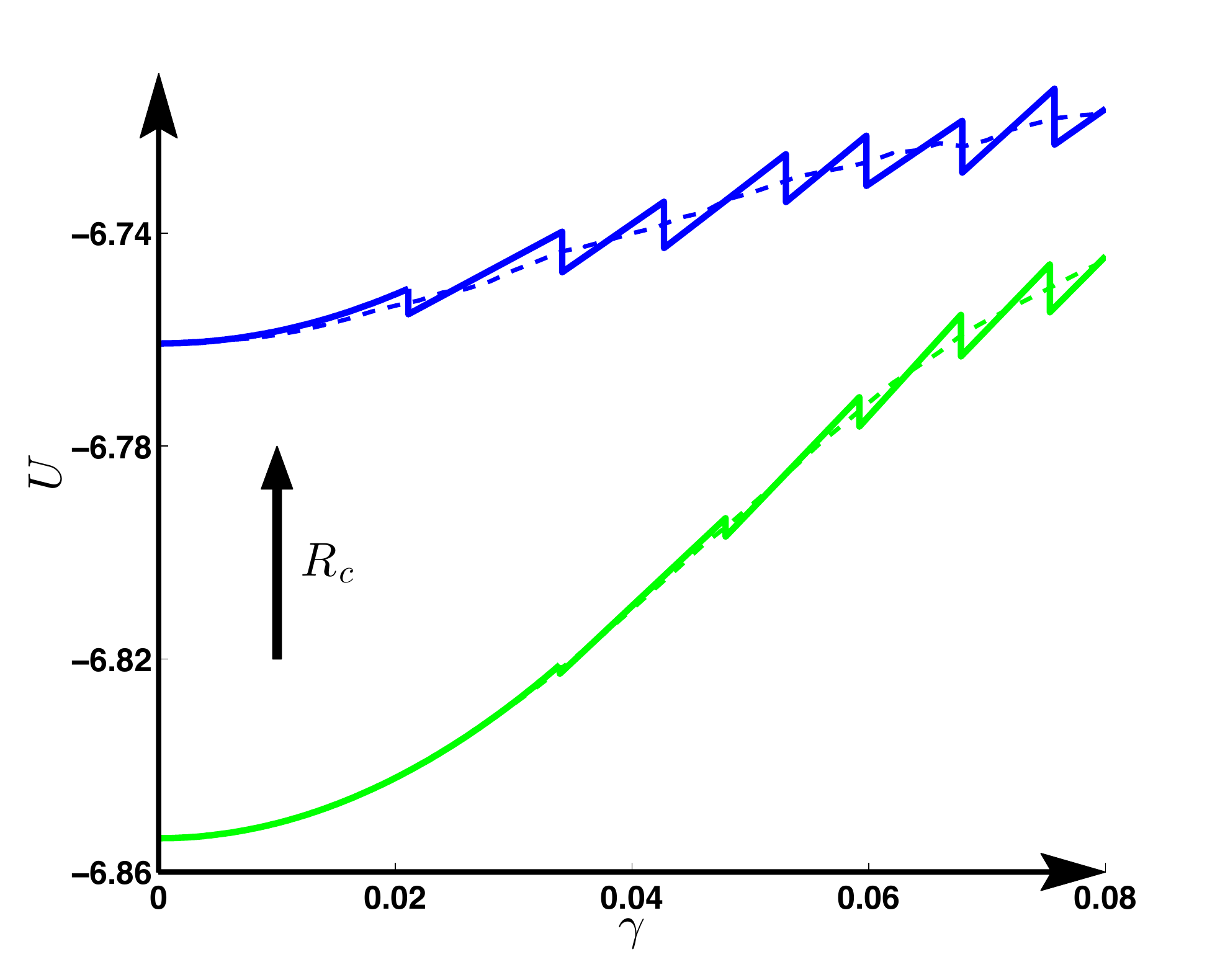}
\caption{Total potential energy per particle $U$ versus strain
$\gamma$ for binary Lennard-Jones glasses containing $N=250$
particles prepared using cooling rates $R_c=10^{-2}$ (top) and
$10^{-5}$ (bottom) and then subjected to athermal, quasistatic pure
shear. The solid and dashed lines indicate single-configuration
(with average particle rearrangement statistics; see Supplemental
Material) and ensemble-averaged $U(\gamma)$, respectively.}
\label{fig:pel}
\end{center}
\end{figure}

The potential energy landscape, which in the case of pure shear is a
function of the $3N$ particle coordinates and strain $\gamma$, can
provide key insights into the structure of configuration space and
mechanical behavior of glassy
materials~\cite{stillinger1995topographic,debenedetti2001supercooled}.
For example, recent studies have identified hierarchical sub- and
metabasins in the energy landscape of supercooled liquids and
glasses~\cite{debenedetti2001supercooled,harmon2007anelastic,charbonneau2014fractal}
and studied the disappearance of energy minima and resulting particle
rearrangements as a function of applied
strain~\cite{malandro1999relationships,lacks2004energy,debenedetti2001supercooled}. In
Fig.~\ref{fig:pel}, we show the total potential energy per particle
$U$ versus strain $\gamma$ for a single configuration (with average
particle rearrangement statistics), as well as $U(\gamma)$ averaged over $500$
configurations, for two cooling rates ($R_c = 10^{-5}$ and
$10^{-2}$). For the more rapidly cooled glass, the energy drops are
relatively large and frequent.  Rearrangements are frequent because
there is an abundance of nearby minima in the potential energy
landscape as a function of strain for rapidly cooled
glasses~\cite{stillinger1995topographic}. In addition, the large and
frequent energy drops give rise to small metabasin curvature of the
ensemble-averaged energy landscape.

In contrast, for slowly cooled glasses, we find that the energy drops
are smaller and more rare, which implies that
when systems are deeper in the energy landscape, energy minima are
further separated as a function of strain and the curvatures of the
energy metabasins are larger. Thus, these results provide evidence
that the curvatures of the metabasins in the potential energy
landscape decrease with increasing cooling
rate~\cite{utz2000atomistic,fan2014thermally}.  In addition, our
results are consistent with prior results, which show that the yield
stress (or the ensemble-averaged $dU/d\gamma$) increases with
decreasing cooling rate~\cite{utz2000atomistic,ashwin2013cooling}.

\begin{figure}
\begin{center}
\includegraphics[width=0.9\columnwidth]{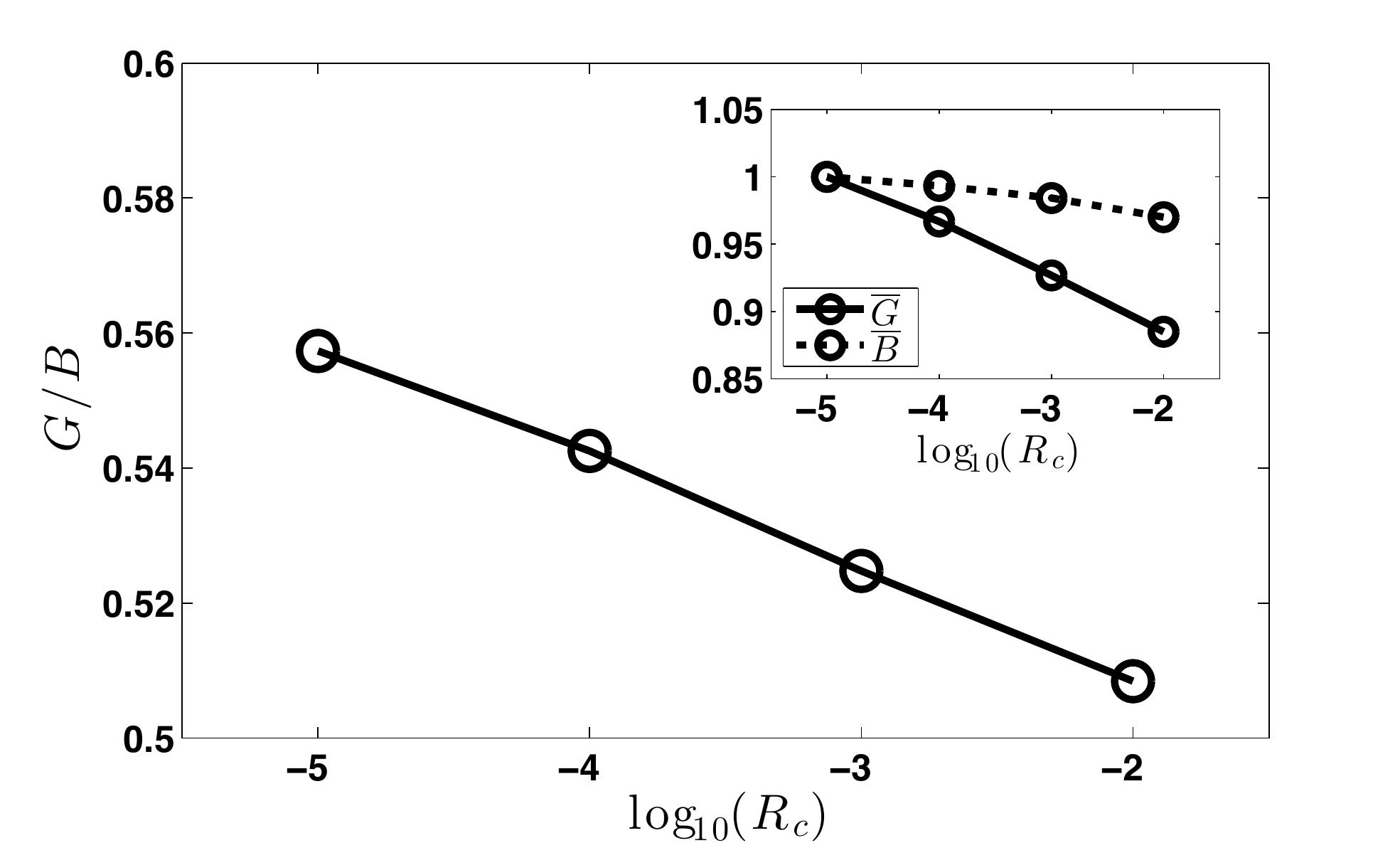}
\caption{The ratio of the shear to bulk modulus $G/B$ as a
function of the cooling rate $R_c$ used to prepare the binary
Lennard-Jones glasses. The inset shows the shear $\overline{G}$ and
bulk moduli $\overline{B}$ separately as a function of $R_c$, both
of which have been normalized to $1$ at $R_c=10^{-5}$.}
\label{fig:GB}
\end{center}
\end{figure}

We next connect the statistics of the potential energy landscape to
micro- and macro-scale mechanical properties of amorphous solids. As
shown in Fig.~\ref{fig:pel}, more rapidly cooled glasses undergo
larger and more frequent particle rearrangements.  We argue that
larger and more frequent particle rearrangements help reduce stress
accumulation during deformation and hence prevent catastrophic brittle
failure~\cite{lewandowski2005intrinsic,zemp2015crystal,yu2012tensile}.
This suggests that more rapidly quenched glasses are more ductile than
slowly quenched glasses~\cite{kumar2013critical}.  To investigate this
hypothesis, we measured the ratio of the shear to bulk modulus $G/B$
(Fig.~\ref{fig:GB}) as a function of cooling rate. (The moduli $G$ and
$B$ were obtained from the slope of $\sigma(\gamma)$ for vanishingly
small pure and compressive strains, respectively.)  $G/B$ is a
material property that has been shown to correlate strongly with the
ductility/brittleness of a
material~\cite{lewandowski2005intrinsic,schroers2004ductile,kumar2013critical,shi2014intrinsic}. We
find that both $G$ and $B$ decrease with increasing cooling rate, but
$G$ decreases faster (inset to Fig.~\ref{fig:GB}), and thus the ratio
$G/B$, and brittleness, decrease with increasing $R_c$.

\begin{figure}
\begin{center}
\includegraphics[width=1.05\columnwidth]{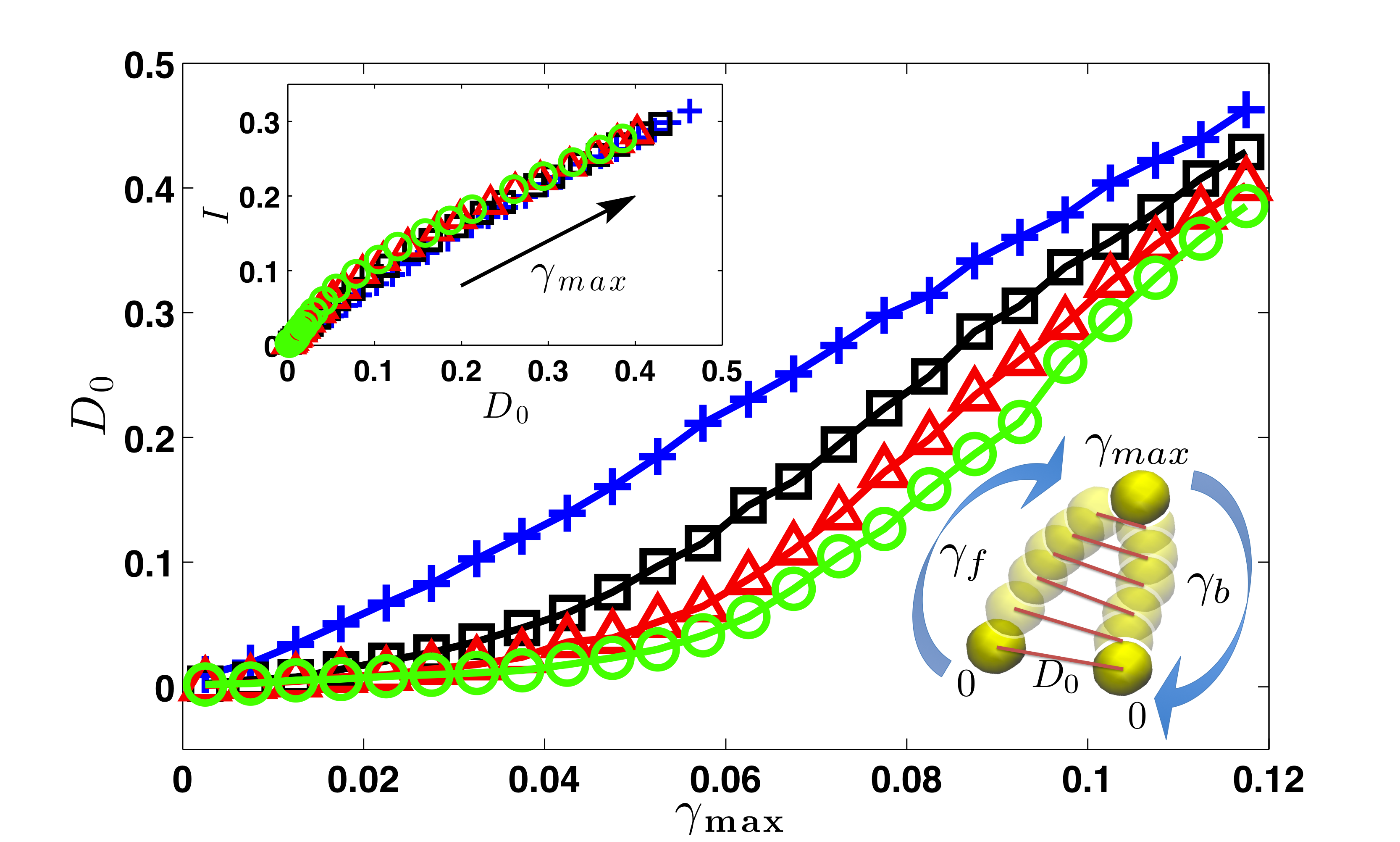}
\caption{Measurement $D_0$ of the degree to which a system deviates from
 the original unsheared configuration after undergoing a single athermal 
quasistatic pure shear cycle with strain amplitude $\gamma_{\rm max}$ for several 
cooling rates: $R_c=10^{-2}$ (crosses), $10^{-3}$ (squares), $10^{-4}$ 
(triangles), and $10^{-5}$ (circles). The curves were averaged over 
$96$ samples with $N=2000$. The upper left inset shows that the two measures 
of irreversibility, $D_0$ and $I$, are positively correlated.  The lower 
right inset shows a schematic of the trajectory of a single particle during 
forward shear from total strain $0$ to $\gamma_{\rm max}$ and backward 
shear from total strain $\gamma_{\rm max}$ to $0$. The bars connect the 
particle positions at corresponding strains during the trajectory, 
${\vec R}(\gamma',0)$ and ${\vec R}(\gamma_{max},\gamma_{max}-\gamma')$.  
$D_0$ and $I$ are related to the length 
of the lower bar and the average length over all bars, respectively.}
\label{fig:irre}
\end{center}
\end{figure}

Whether a material is reversible or not under applied deformation is
often inferred from the behavior of its stress-strain curve or other
macroscopic measurements.  For example, materials are typically deemed
reversible in the regime where the stress-strain curve is linear, and
irreversible in the regime where plastic flow
occurs~\cite{taub1981ideal}.  Reversibility has been studied
experimentally using enthalpy~\cite{harmon2007anelastic} and strain
recovery~\cite{fujita2012low}, elastostatic
compression~\cite{park2008elastostatically},
nanoindentation~\cite{Atzmon_2014}, and quality factor
measurements~\cite{kanik2014high}.  In simulations, reversibility has
been studied using cyclic shear of model
glasses~\cite{schreck2013particle,regev2013onset,regev2015reversibility,
  fiocco2014encoding,priezjev2016reversible}.  Though the large
elastic region ({\it e.g.} the linear stress-strain region in
Fig.~\ref{fig:event}) is typically considered reversible, recent
measurements have been identified irreversible events and anelasticity
on the micro-scale in this `elastic'
region~\cite{park2008elastostatically,fujita2012low,dmowski2010elastic,schall2007structural,jensen2014local,ju2011atomically,ding2012correlating}.

An important, fundamental open challenge is to determine the onset~\cite{schreck2013particle,regev2013onset} of
micro-scale irreversibility and understand its connection to
irreversibility and plasticity on macroscopic scales.  Above, we
defined particle rearrangements as those that led to local
irreversibility of the potential energy after a forward strain
increment $d\gamma$, followed by a backward strain increment
$-d\gamma$.  We now characterize the reversibility of particle motion
following {\it finite-sized} strains using two measures.  First, we 
define a measure of ``state'' irreversibility as
\begin{equation}
\label{d}
D_0(\gamma) = |{\vec R}(0,0)-{\vec R}(\gamma,\gamma)|/N,
\end{equation}
where ${\vec R}(\gamma_f,\gamma_b)$ gives the particle coordinates
after the system has been sheared forward by strain $\gamma_f$ and backward
by strain $\gamma_b$. $D_0$ characterizes the ability of a sheared system to
return to the original, unsheared configuration. (See the lower right inset to
Fig.~\ref{fig:irre}.) In Eq.~\ref{d}, ${\vec R}(0,0)$ gives the
original, unsheared particle coordinates, and ${\vec
  R}(\gamma,\gamma)$ gives the particle coordinates of the system
after it was sheared forward to strain $\gamma$ and then sheared
backward from strain $\gamma$ to zero strain.  During forward strain, the system undergoes
$N_r(\gamma)$ particle rearrangements and during backward strain, the
system undergoes a different set of $N_r(-\gamma)$ particle
rearrangements.  $D_0(\gamma) \sim 0$ indicates a type of reversible behavior,
where most of the particles return to their original, unsheared
positions after a strain cycle of amplitude $\gamma$. In contrast, $D_0 > 0$ 
implies irreversible behavior that grows in magnitude with increasing $D_0$. 
We also define a measure of ``path''
irreversibility,
\begin{equation}
I(\gamma)=\frac{1}{N}\sqrt {\frac{1}{\gamma}\int_0^{\gamma} |{\vec R}(\gamma',0)-{\vec R}(\gamma,\gamma-\gamma')|^2 d\gamma'},
\end{equation}
which determines the average distance between the system
configurations at corresponding strains during the forward and
backward portions of the shear cycle.  
(See the lower right inset to Fig.~\ref{fig:irre}.)

Even though shear cycles can occur with $I>0$ and $D_0=0$, which
implies that the system returns to the original, unsheared
configuration at $\gamma=0$ along different forward and backward shear
paths~\cite{schreck2013particle}, we find that the ensemble-averaged
$I$ becomes nonzero only when $D_0$ begins increasing from zero.
Further, $I$ and $D_0$ are strongly correlated as the amplitude
$\gamma_{\rm max}$ of the shear cycle increases. (See the upper left
inset in Fig.~\ref{fig:irre}.) In Fig.~\ref{fig:irre}, we plot
$D_0(\gamma_{max})$ for several cooling rates.  We find that slowly
cooled glasses are nearly reversible over a finite range of strain
({\it i.e.} up to $\gamma_{\rm max} \sim \gamma_y = 0.05$), while $D_0
\sim A \gamma_{\rm max}$ (with slope $A$) for rapidly cooled
glasses. For intermediate cooling rates, $D_0$ can be approximated as
$D_0 \sim B(R_c) \gamma_{\rm max}$ for $\gamma_{\rm max} <
\gamma_y(R_c)$ and $D_0 \sim A \gamma_{\rm max}$ for $\gamma_{\rm max}
> \gamma_y(R_c)$. The slope $B(R_c)$ increases with cooling rate, and
the crossover strain $\gamma_{y}(R_c)$ decreases with cooling rate.
The upper inset to Fig.~\ref{fig:irre} shows that $I$ and $D_0$
possess the same cooling rate dependence.  More rapidly cooled glasses
possess higher values of irreversibility because they undergo more
frequent and larger rearrangements during shear (as shown in
Fig.~\ref{fig:pel}).  In addition, we have shown that the path
irreversibility $I$ is strongly correlated with the energy loss per
rearrangement $dU_{\rm loss}/dN_r$.

It is well known that non-affine particle motion and rearrangements
control the mechanical properties of amorphous solids.  In this study,
we showed that the particle rearrangement statistics are sensitive to
the cooling rate used to prepare zero-temperature
glasses. Specifically, we identified distinct particle rearrangement
events in binary Lennard-Jones glasses undergoing athermal,
quasistatic pure shear.  We measured the frequency of rearrangements
and the size of the energy drops during rearrangements as a function
of strain and the cooling rate used to prepare the glasses. We found
that more rapidly cooled glasses undergo more frequent and larger
energy drops, compared to more slowly cooled glasses. We also
correlated the statistics of particle rearrangements to the ratio of
the shear and bulk moduli and showed that more rapidly cooled glasses
are more ductile than slowly cooled glasses. Finally, we characterized
the degree of irreversibility of systems to cyclic shear and showed
that slowly cooled glasses possess a finite range of strain where they
appear nearly reversible (but are not truly reversible with
$D_0=I=0$), whereas measures of irreversibility increase linearly with
strain for rapidly cooled glasses. Thus, we showed that more ductile
glasses are also more irreversible, and connected particle-scale 
rearrangement statistics to macroscopic mechanical response. 

\begin{acknowledgments}
The authors acknowledge primary financial support from NSF MRSEC
DMR-1119826 (K.Z.) and partial support from NSF Grant Nos.
CMMI-1462439 (C.O. and M.F.) and CMMI-1463455 (M.S.). 
This work also benefited from
the facilities and staff of the Yale University Faculty of Arts and
Sciences High Performance Computing Center and the NSF (Grant
No. CNS-0821132) that, in part, funded acquisition of the computational
facilities.
\end{acknowledgments}

\bibliography{reversibility_PRL}

\end{document}